\documentclass[%
 aip,
 rsi,
 amsmath,amssymb,
  reprint%
]{revtex4-1}

\usepackage{graphicx}
\usepackage{dcolumn}
\usepackage{bm}

\usepackage[utf8]{inputenc}
\usepackage[T1]{fontenc}
\usepackage{mathptmx}
\usepackage{braket}

\begin{document}

\preprint{AIP/123-QED}

\title[]{Long distance optical transport of ultracold atoms: \\A compact setup using a Moir\'e lens}

\author{G. Unnikrishnan}%
 \author{C. Beulenkamp}%
 \author{D. Zhang}%
 \author{K.P. Zamarski}%
 \author{M. Landini}%
 \author{H.-C. N\"agerl}%
  \email{christoph.naegerl@uibk.ac.at}
\affiliation{ 
Institut f{\"u}r Experimentalphysik und Zentrum f{\"u}r Quantenphysik, Universit{\"a}t Innsbruck, 6020 Innsbruck, Austria
}%


\begin{abstract}
We present a compact and robust setup to optically transport ultracold atoms over long distances. Using a focus-tunable Moir\'e lens that has recently appeared on the market, we demonstrate transport of up to a distance of 465 mm. A transfer efficiency of 70\%  is achieved with negligible temperature change at 11 $\mu$K. With its high thermal stability and low astigmatism, the Moir\'e lens is superior to fluid-based varifocal lenses. It is much more compact and stable than a lens mounted on a linear translation stage, allowing for simplified experimental setups.  
\end{abstract}
\maketitle

\section{INTRODUCTION}
Research in ultracold atoms has progressed from the production of atomic Bose-Einstein condensates and degenerate Fermi gases to richer physical systems such as ultracold dipolar molecules and magnetic dipolar atoms.\cite{Anderson198,HuletBEC,KetterleBEC,FermiGas,PolarMoleculesReview2017,PfauCrBEC} Ultracold gases are promising candidates to carry out quantum simulations of condensed matter systems,\cite{QuantSimul_GrossandBloch2017,QuantSimul_Bloch2012} precision measurements of fundamental constants,\cite{MoleculesReview,OpticalClockReview} quantum computing, and quantum sensing.\cite{QcomputingAtomsReview,CappellaroQSensing} Until recently, such systems have been mostly probed via measurements of the bulk properties of these systems, such as temperature and momentum distributions. However, recent work has demonstrated direct measurements of microscopic properties by high resolution imaging.\cite{Bakr2009_GQM,Sherson2010_QGM,Yamamoto_QGM,Cheuk_QGM,Haller2015_QGM,Edge_QGM,Parson_QGM,Omran_QGM} For obtaining a high numerical aperture, this technique often requires optics to be placed very near the sample to be probed. This requirement is usually met by having two vacuum chambers, one for the initial collection of atoms using a magneto-optic trap (MOT) and a second chamber with exquisite vacuum conditions, a clean electro-magnetic environment, and high optical access, necessitating a transport to the second chamber. Apart from high resolution imaging applications, transport of cold atoms is crucial in many experiments aiming to couple ultracold atoms to other quantum mechanical systems.\cite{EsslingerQED,HanschOptomechanics,NanofiberRauschenbeutel}

 Atom transport has been achieved by both magnetic and optical methods. For magnetic transport, the coil system producing the magnetic trap is either moved using a translation stage\cite{MagneticTransportMovingCoil}, or a series of partially overlapping coils is used.\cite{MagneticTransportCoilSet} Apart from being bulky and reducing optical access, a limitation of magnetic transport is that it is restricted to atoms in a quantum state that is both trappable and stable in a magnetic trap, making it unsuitable for atoms like e.g. Cs.\cite{Weber232,Kraemer2004} 

Alternatively, optical transport can be carried out by using a moving optical lattice\cite{LatticeTransportKuhr278}, a movable dipole trap\cite{GRIMM200095} generated by a lens mounted on a linear translation stage\cite{KetterleOpticalTransport}, or by a varifocal lens.\cite{Leonard_2014} Optical lattices are convenient for short distance transport (few mm), although relatively long-distance transport using a Bessel beam has been demonstrated.\cite{Schmid_2006} 
Moving a lens requires a high quality (low jerk, high accuracy and repeatability) stage, which is expensive and bulky. Furthermore, large moving parts are undesirable near experiments that are sensitive to vibrations and magnetic field variations. Recently, fluid based varifocal lenses with small moving parts have been used as a compact system for optical transport\cite{Leonard_2014}. However,  these lenses are sensitive to thermal fluctuations, necessitating additional stabilization mechanisms. Some of these lenses show strong astigmatism, severely affecting the confinement strength of a dipole trap produced using such a lens.

\section{The Moir\'e lens based transport scheme}

Here, we present a transport scheme using a varifocal Moir\'e lens.\cite{MoireExp,MoireTheory} In contrast to fluid-based designs, Moir\'e lenses can be fabricated using fused silica. This makes the lens robust against thermal lensing at high optical powers, rendering further stabilization mechanisms unnecessary. Secondly, astigmatism is low and the setup is very compact and relatively cheap in comparison to a system with an air-bearing linear translation stage. However, one limitation is the fact that only about 53\% of the light incident on the Moir\'e lens is useful, as discussed later.     

\begin{figure}
    \includegraphics[width=0.45\textwidth]{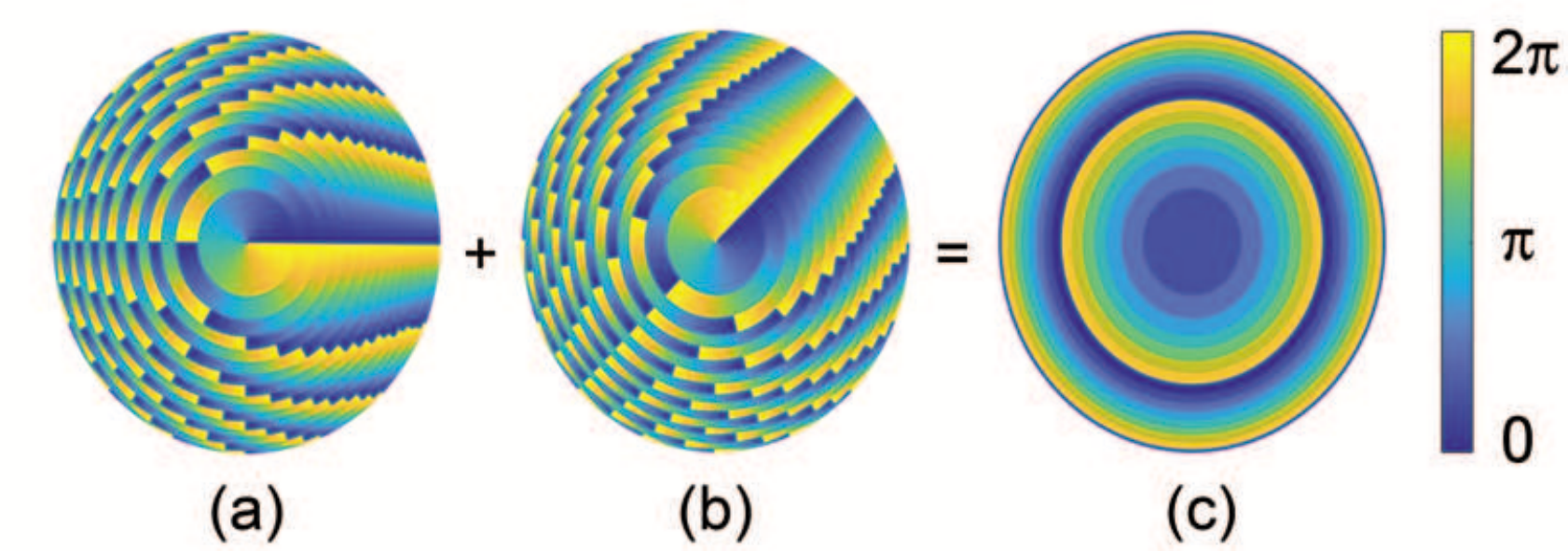}
    
     \caption{An illustration of the phase profile formed by two phase plates as described in Ref. 33 and 34. The second phase plate (b) is rotated by 45 degrees with respect to the first phase plate (a). The resultant phase profile (c) is equivalent to that of a Fresnel lens.}
     \label{fig:phase profile}
\end{figure}

\begin{figure}
    \includegraphics[width=0.45\textwidth]{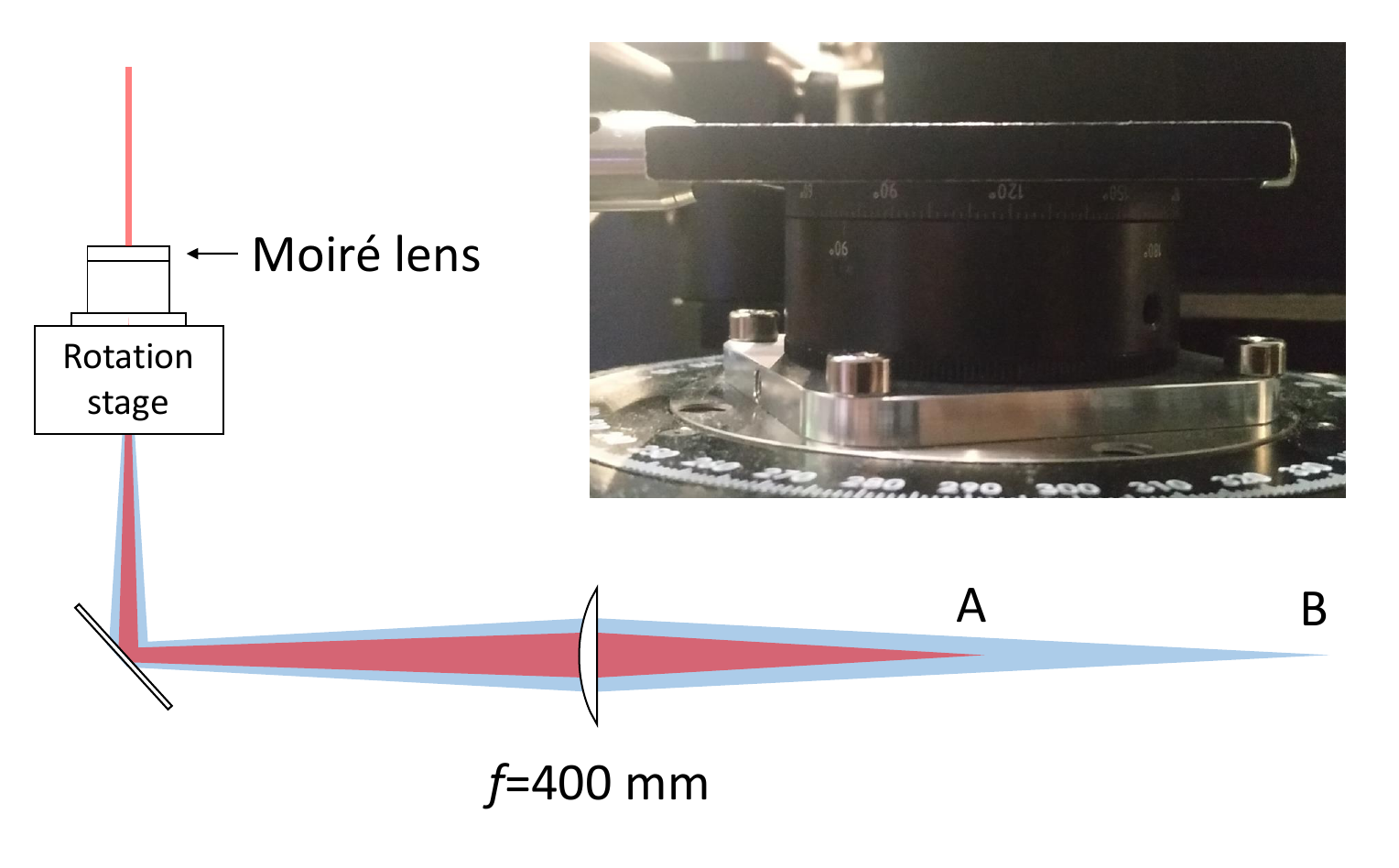}
    \caption{Optical setup. The rotation stage is mounted horizontally on a breadboard. The Moir\'e lens is attached to a home-built aluminium adapter that is bolted onto the rotation stage. The other side of the lens is kept fixed by a clamp. The beam path length between the Moir\'e lens and the static lens is 400 mm. After passing through a static lens mounted on a mirror mount, the beam is sent to the vacuum chamber using steering mirrors. This setup ensures that the beam waist throughout the transport range from A to B is constant. The inset shows a photograph of the lens mounted on the rotation stage.}
    \label{fig:optical setup}
\end{figure}

A Moiré lens consists of two phase plates imprinting a specific phase pattern on the incident wavefront. Such a pattern is etched onto the plates with a finite resolution (pixel size) as shown schematically in Fig. \ref{fig:phase profile}. When two such phase plates are combined, a radial phase profile corresponding to that of a lens is formed, as illustrated in Fig. \ref{fig:phase profile}. By rotating the plates with respect to each other, the focal length of the lens can be tuned. The optical power $D$ of the Moir\'e lens used in our setup is given by:
\begin{equation}
    D=\frac{4 \theta \lambda}{A s\sqrt{4 \pi^2+1} }, 
\end{equation}
where $\theta$ is the angle between the two phase plates in radians, $\lambda$ is the wavelength of light, $A=25$ mm is the aperture of the lens, and $s=8 \mu$m is the pixel size on the phase plates. The lens is sold as a single unit with the two phase plates pre-mounted on a rotation mount, as shown in the inset of Fig. \ref{fig:optical setup}.

\begin{figure}
   \includegraphics[width=0.45\textwidth]{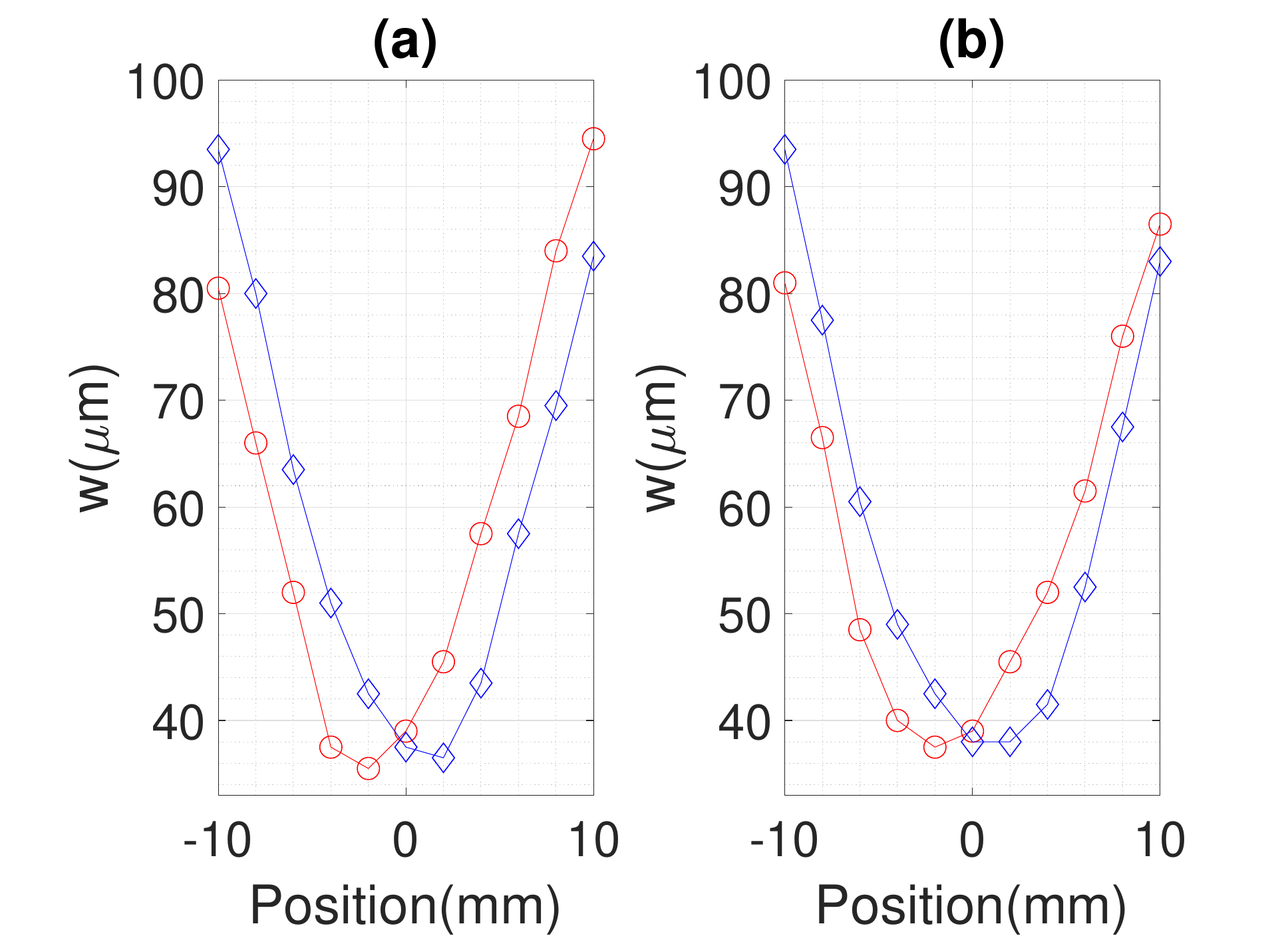}
 
    \caption{Beam waist measurements to characterize the astigmatism of the Moir\'e lens. The 1/e$^2$-waists along two perpendicular axes obtained from a Gaussian fit are plotted as a function of distance. (a) At the initial position of transport (b) 232.5 mm from the initial position. Error bars are smaller than the symbol size.}
    \label{fig:beam profiles}
\end{figure}

In our transport scheme, an optical dipole trap produced by a focused laser beam is used to trap and move an atomic sample from an ultra-high vacuum stainless-steel chamber towards a quartz cell with high optical access. The focused laser beam is produced using the setup presented in Fig \ref{fig:optical setup}. The trap position is changed by varying the focal length of the varifocal Moir\'e lens. Since the depth $U$ of such a trap is proportional to $P/w_0^2$, where $P$ is the power in the laser beam and $w_0$ is the waist of the Gaussian beam at the focus, $w_0$ needs to be constant throughout the transport range to maintain constant trapping conditions. For this, we adopt the technique presented in L\'eonard {\it et al.}\cite{Leonard_2014}, whereby a second static lens is placed at a distance such that the varifocal lens is in its Fourier plane. 
Up to 10 W of 1064-nm light is delivered via a photonic crystal fiber (NKT, aeroGUIDE-5-PM), after which the beam is enlarged using a 5x telescope to a 1/e$^2$-waist of $\sim$3.4 mm. The beam then passes through the Moir\'e lens with an aperture of 25 mm (MIMF-25-1064 from Diffratec) and a static lens with $f=$400 mm, producing a constant 1/e$^2$-waist of $\sim$38 $\mu $m throughout the transport range of 232.5 mm. Both lenses are anti-reflection coated for 1064 nm. As shown in Fig \ref{fig:optical setup}, one side of the Moir\'e lens is attached onto a home-built adapter that is bolted to a rotation stage, while the other side is clamped externally. The beam qualities near the focal point at two different positions are shown in Fig \ref{fig:beam profiles}. We note that the separation between the foci is small and that the beam waists at the two positions are close to identical. 

For producing a relative rotation between the two phase plates in the Moir\'e lens, we use a mechanically driven rotation stage (Newport RGV100BL-S), with dimensions of 115 mm $\times$ 115 mm $\times$ 60 mm. 
For every degree of relative rotation, the focus position changes linearly by 10.47 mm. 
According to the manufacturer, the stage provides an accuracy of 5 millidegrees, 0.1 millidegree minimum incremental motion, a maximum angular acceleration of 1000 deg/s$^2$, and a maximum angular velocity of 720 deg/s,  corresponding to 52 $\mu $m accuracy, 1 $\mu $m minimum incremental motion, 10473 mm/s$^2$ acceleration, and 7541 mm/s velocity, respectively.

\section{The experiment and results}

Our experiment begins with a three-dimensional MOT of $^{39}$K atoms loaded from a 2D$^+$ MOT.\cite{KCs-1} After sub-Doppler cooling using a grey molasses technique, the atoms are loaded into a tightly confining magnetic trap and subsequently into a very deep dipole trap. From this trap, atoms are transferred into the focus of the transport beam. 
At this point, we have up to 5$\times$10$^5$ atoms in the state $\ket{F=1,m_F=-1}$ at a temperature of 11 $\mu$K with 3.1 W of optical power on the Moir\'e lens. By inducing oscillations along the axial direction of the trap, we measure a trap frequency of $\omega_\text{axial}/ 2\pi=$4.42(5) Hz. The radial trapping frequency is measured via parametric heating to be $\omega_\text{radial}/ 2\pi=$972(3) Hz. The estimated trap depth is $U/ k_\text{B}=$87 $\mu$K, where $k_\text{B}$ is the Boltzmann constant.
\begin{figure}
    \includegraphics[width=0.45\textwidth]{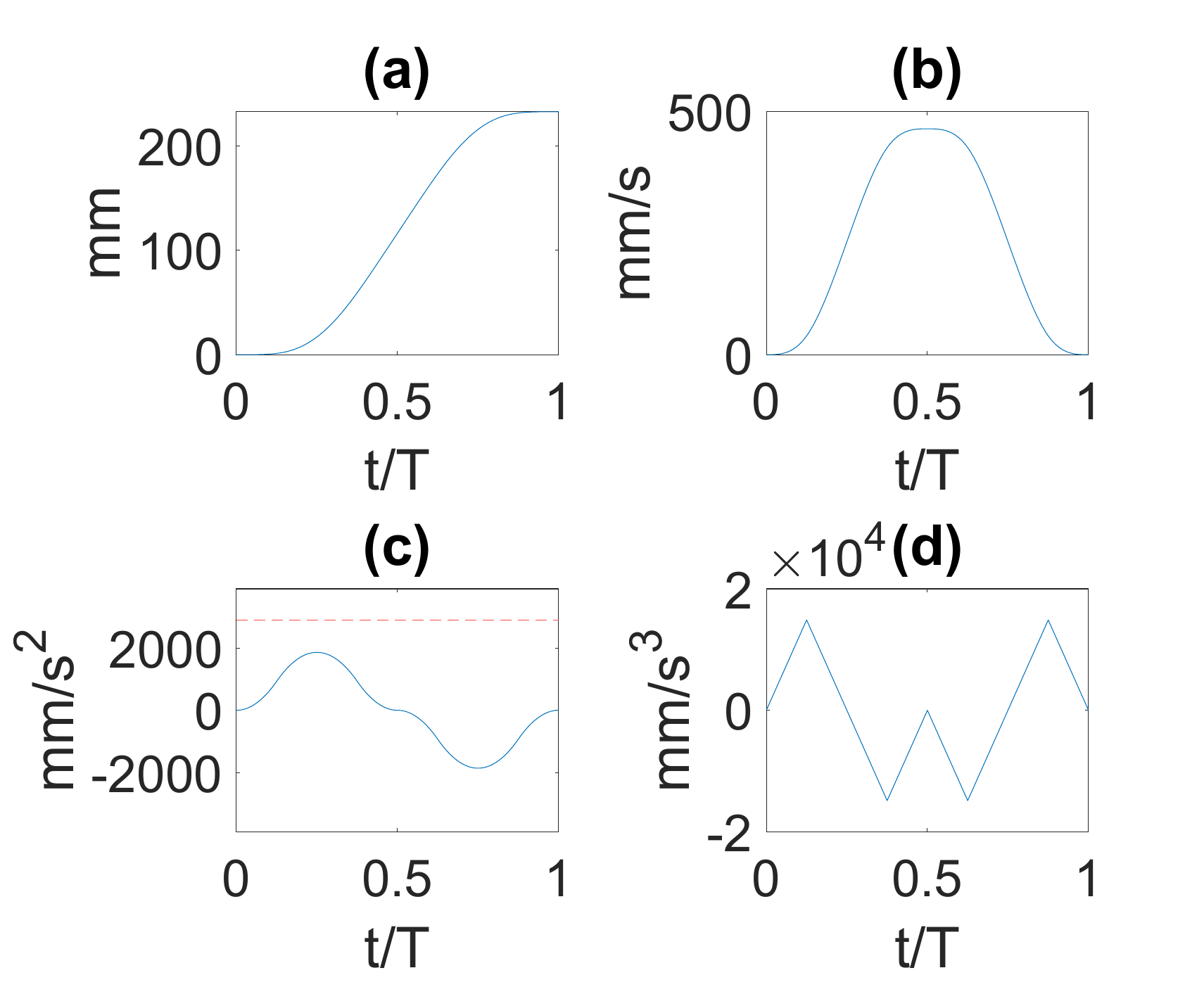}
    \caption{Motion profiles used for one-way transport by varying the relative angle $\theta$ by 22.2 degrees. (a) Displacement from the initial position, (b) velocity, (c) acceleration, and (d) jerk as a function of time $t$ in units of the total transport time $T$. The red dashed line in (c) indicates the value of $a_\text{max}$ for 3.1 W power on the Moir\'e lens.}
    \label{fig:motion profiles}
\end{figure}

\begin{figure}
    \includegraphics[width=0.5\textwidth]{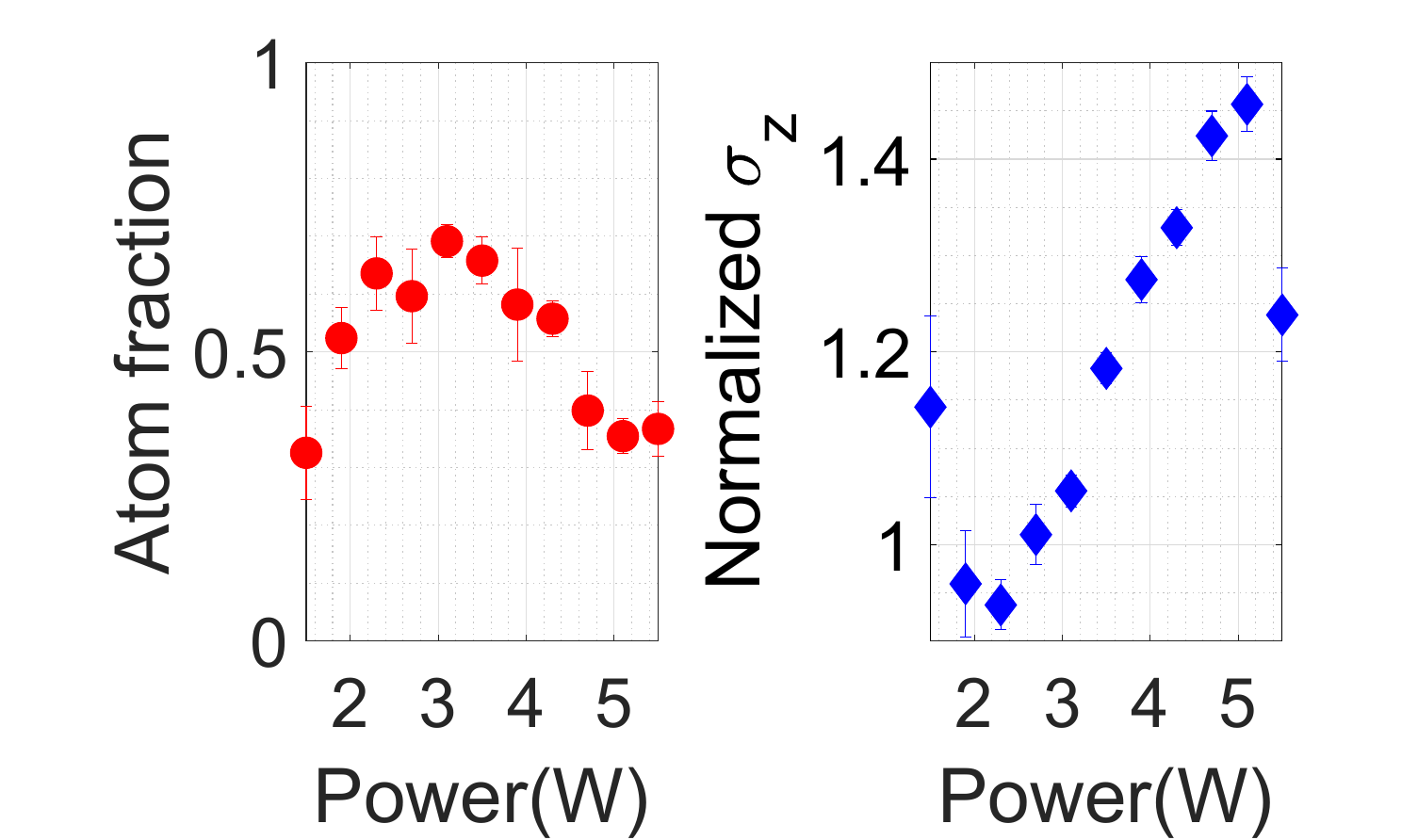}
    \caption{Atom fraction and heating for 3.2 s roundtrip transport time. The red circles indicate the atom number after transport normalized to the atom number after holding in the trap for 3.2 s.  The blue diamonds denote the radial cloud size $\sigma$ after transport normalized to the cloud size after holding in the trap for 3.2 s.  }
    \label{fig:transport efficiency}
\end{figure}

To ensure a smooth transport of the atomic sample, we apply the in-built S-shaped ramp for the angle provided by the stage's motion controller (Newport XPS-RLD). The motion profiles for which we found optimal transport efficiency are shown in Fig. \ref{fig:motion profiles}. To quantify the efficiency of the transport, we move the atomic cloud to a fixed distance of 232.5 mm (which is half the distance to the quartz cell, see the discussion below) and then back to the original position (roundtrip 465 mm), where we image the sample to determine the number of remaining atoms and the temperature of the atomic cloud. Our results are shown in Fig \ref{fig:transport efficiency} for various optical powers for a total transport time of 3.2 s. With increasing laser power, the transfer fraction increases and reaches a peak at about 3.1 W before dropping off again. At 3.1 W, the roundtrip transfer efficiency is 70\%. The temperature of the cloud, as quantified by a normalized cloud size, shows a non-monotonous dependence on the laser power: after an initial drop, it rises before dropping again. At the peak transfer efficiency of 3.1 W, the temperature is found to essentially remain constant. 
A temperature measurement at this point gives 10 $\mu$K, which is about the same as the initial temperature and also the same as for a non-transported sample held for the duration of the transport. Note that over a timescale of 3.2 s, we find about 15\% atom loss and negligible temperature increase for a non-transported sample just by holding the sample in the trap. We attribute the additional loss of atoms without temperature increase during the transport to evaporative cooling counteracting heating due to partial non-adiabaticity of the transport. We define adiabaticity as $a_\text{max}<<F_\text{axial}/m_\text{K}$, where $a_\text{max}$ is the maximum acceleration during transport, $m_\text{K}$ the mass of $^{39}$K and $F_\text{axial}$ is the maximum trapping force of the dipole trap along the axial direction. As shown in Fig. \ref{fig:motion profiles} (c), the maximum acceleration reached during transport is not negligible in comparison to the maximum allowed acceleration $a_{max}$. 

\section{Discussion and summary}
In addition to being very compact, optical transport using the Moir\'e lens is less prone to vibrations when compared with a linear translation stage. For a linear translation stage, while vibrations along the transport direction are minimized using a good feedback controller, vibrations perpendicular to the axis of the motion induce pointing deviations that transfer energy to the atoms. Energy transferred to the atomic sample is proportional to the strength of the power spectral density of vibrations at twice the trap frequency.\cite{LaserNoiseHeating} We note that typical axial and radial trap frequencies in a dipole trap are in the range of a few Hz and a few kHz, respectively. Since typical vibrational noise sources are likely to be in the kHz range, this makes noise that couples to the radial direction more relevant. An analogous situation producing pointing deviations in the case of a Moir\'e lens would be tilts of the rotation axis, which is less likely to occur since the lens is fixed to the rotation stage. This makes optical transport using such a setup a very robust scheme.

Another feature of the Moir\'e lens that needs to be taken into consideration is its finite focusing efficiency $\eta$. As described in Bernet {\it et al.}\cite{MoireExp,MoireTheory}, $\eta$, defined as the amount of power in the focal spot divided by the total power input to the lens, varies with the relative angle between the phase plates of the lens as sinc$^2(\theta/2)$ and has a maximum value of $\sim$53\%. The peak value is limited to $53\%$ mainly as a result of the finite size of the pixels used in the phase plates to realize the continuous phase pattern, leading to half of the power being lost in other diffraction orders. In practice, the other diffraction orders are co-propagating with the used beam but with foci far away from the used order. All orders are eventually sent into a beam dump. Although the peak efficiency is limited, the variation in efficiency when varying $\theta$ can be counterbalanced by e.g. acousto-optic techniques.
In our setup, we vary $\theta$ from -14.7 degrees to -36.9 degrees, resulting in an $\eta$ variation of 1.5\%, which enables transport even without compensating for the change in $\eta$. We note that even larger transport ranges can be achieved with acceptable changes in $\eta$ by shifting the required focal range to lie around $\theta=0$  by utilizing an additional offset lens. In fact, we transported our samples up to 465 mm in one direction, i.e. into the quartz cell. However, insufficient vacuum conditions in the quartz cell turned out to limit the transport efficiency. Furthermore, an even smoother motion profile can potentially improve the transport efficiency.

To summarize, we have demonstrated transport  of ultracold atoms over a large distance of 465 mm using a varifocal Moir\'e lens with up to 70\% transfer efficiency and negligible change in the temperature of the atomic cloud. In comparison to previous techniques, our setup is very compact, less prone to vibrations, and uses a relatively cheap mechanically driven rotation stage instead of an expensive air-bearing stage. Secondly, the lens material, fused silica, has excellent thermal properties at high optical powers needed for trapping atoms, making it superior to fluid-based lenses.  Our method will be useful for experiments requiring a simple, robust, and compact optical setup for transporting cold atoms.

\section{Acknowledgements}

We thank A. Wang for an initial characterization of the Moir\'e lens. G. U. and K.P.Z acknowledge support by the Austrian Science Fund (FWF) within the DK-ALM (grant no. W1259-N27). We acknowledge funding by the European Research Council (ERC) under project no. 789017, by the FWF under project no. P29602-N36, and by a Wittgenstein prize grant under FWF project no. Z336-N36.

\section*{DATA AVAILABILITY}
The data that support the findings of this study are available from the corresponding author upon reasonable request.

\nocite{*}

\bibliography{aipsamp}

\end{document}